\documentclass[12pt,a4paper,preprint]{article}

\usepackage{graphicx}
\usepackage{amsmath}
\usepackage{amssymb}
\usepackage{amsxtra}
\usepackage{multicol}
\usepackage{subfigure}
\usepackage{geometry}

\def\Journal#1#2#3#4{{#1} {\bf #2} (#4) #3}
\def\PLB{{ Phys. Lett.}  B}
\def\PRL{ Phys. Rev. Lett.}
\def\PRD{{ Phys. Rev.} D}

\def\GaC{ Gravitation and Cosmology}

\def\JETPL{ JETP Lett.}

\def\CQG{ Class. Quantum Grav.}
\def\APJ{ Astrophys. J.}

\def\IJMPA{{ Int. J. Mod. Phys.}  A}

\def\BWP{ Bled Workshops in Physics}

\def\s{{\,\rm s}}

\def\GeV{\,{\rm GeV}}

\def\({\left(}
\def\){\right)}

\def\kpc{{\,\rm kpc}}

\graphicspath{{figDMdecays/}}
\title{High Energy Positrons and Gamma Radiation from Decaying Constituents of a
two-component Dark Atom Model}
\author{K.~Belotsky$^{1,2}$, M.~Khlopov$^{1,2,3}$, C. Kouvaris$^{4}$, M.~Laletin$^{1}$\\$^{1}$National Research Nuclear University MEPHI\\(Moscow Engineering Physics Institute),\\ Moscow, Russia \\$^{2}$ Centre for Cosmoparticle Physics ``Cosmion"\\ Moscow, Russia \\$^{3}$ APC laboratory 10, rue Alice Domon et L\'eonie Duquet \\75205 Paris Cedex 13, France\\$^{4}$$\text{CP}^3$-Origins, University of Southern Denmark,\\ Campusvej 55, Odense 5230, Denmark}
\date{}
\begin{document}
\maketitle
\begin{center}

\end{center}
\begin{abstract}
We study a two component dark matter candidate inspired by the Minimal Walking Technicolor model. Dark matter consists of a dominant SIMP-like dark atom component made of bound states between primordial helium nuclei and a doubly charged technilepton, and a small WIMP-like component made of another dark atom bound state between a doubly charged technibaryon and a technilepton. This scenario is consistent with direct search experimental findings because the dominant SIMP component interacts too strongly to reach the depths of current detectors with sufficient energy to recoil and the WIMP-like component is too small to cause significant amount of events. In this context a metastable technibaryon that decays to $e^+e^+$, $\mu^+ \mu^+$ and $\tau^+ \tau^+$ can in principle explain the observed positron excess by AMS-02 and PAMELA, while being consistent with the photon flux observed by FERMI/LAT. We scan the parameters of the model and we find the best possible fit to the latest experimental data. We find that there is a small range of parameter space that this scenario can be realised under certain conditions regarding the cosmic ray propagation and the final state radiation. This range of parameters fall inside the region where the current run of LHC can probe, and therefore it will soon be possible to either verify or exclude conclusively this model of dark matter.

\end{abstract}
%\preprint{CP3-Origins-2015-033 DNRF90, DIAS-2015-33.}
{\footnotesize \it Preprint: CP3-Origins-2015-033 DNRF90, DIAS-2015-33.}
\section{Introduction}\label{intro}
Dark matter in the form of dark atoms can be a viable solution to the current experimental puzzle. Even stable electrically charged particles can exist hidden in such atoms,  bound by  ordinary Coulomb interactions (see \cite{mpla,DMRev,DDMRev} and references therein).
Stable particles with charge -1 are excluded due to overproduction of anomalous isotopes of hydrogen.  However, for negatively doubly charged particles the constraints are much weaker and this possibility remains open.

There
are several  models where heavy stable -2  charged species $O^{--}$ can emerge naturally:
\begin{itemize}
\item[(a)] AC-leptons, from theories beyond the Standard Model, based on the approach
of almost-commutative geometry \cite{Khlopov:2006dk,5,FKS,bookAC}.
\item[(b)] Technileptons and
anti-technibaryons in the framework of Walking Technicolor
(WTC) \cite{KK,Sannino:2004qp,Hong:2004td,Dietrich:2005jn,Dietrich:2005wk,Gudnason:2006ug,Gudnason:2006yj}.
\item[(c)] stable ``heavy quark clusters" $\bar U \bar U \bar U$ formed by anti-$U$ quark of a 4th generation
\cite{Khlopov:2006dk,Q,I,lom,KPS06,Belotsky:2008se}.
\end{itemize}
All these models 
predict also corresponding +2 charge particles. If these positively charged particles remain free in the early Universe,
they can recombine with ordinary electrons in anomalous helium, which is strongly constrained in
terrestrial matter. Therefore a cosmological scenario should provide a  mechanism which suppresses anomalous helium.
There are three possible mechanisms that can provide a suppression:
\begin{itemize}
\item[(i)] The abundance of anomalous helium in the Galaxy may be significant, but in terrestrial matter
 a recombination mechanism could suppress it below experimental upper limits \cite{Khlopov:2006dk,FKS}.
This possibility requires a new U(1) gauge symmetry which materialises as a  long range interaction among dark matter particles mediated by a dark photon.
%The existence of a new U(1) gauge symmetry, causing new Coulomb-like long range interactions between charged dark matter particles, is crucial for this %mechanism. This leads inevitably to the existence of dark radiation in the form of hidden photons.
\item[(ii)] Free positively charged particles are already suppressed in the early Universe and the abundance
of anomalous helium in the Galaxy is negligible \cite{mpla,I}.
\item[(iii)] Excessive positively charged particles are created in deficit relative to excessive negatively charged particles and bind with them in neutral atom-like states \cite{Bled14,Belotsky:2014}.
\end{itemize}
These three possibilities correspond to three different cosmological scenarios of dark atoms. The first one is
realized in the scenario with AC leptons, forming neutral AC atoms \cite{FKS}.
The second assumes a charge asymmetry  of the $O^{--}$ which forms the atom-like states with
primordial helium \cite{mpla,I}. In the third case, which can be realized in WTC, the excess of positively charged technibarions $UU^{++}$ is much smaller, than the excess of negatively charged $\zeta^{--}$, so that all the free $UU^{++}$ are captured by $\zeta^{--}$ in ($\zeta^{--} UU^{++}$) neutral bound states \cite{Bled14,Belotsky:2014}. In WTC (see~\cite{Gudnason:2006yj} and references therein for a review) 
new stable species (techniquarks and technileptons) that transform under  nontrivial representations of
the SU(2) electroweak group, participate in  sphaleron processes that eventually determine the relative abundance between baryons  and 
-2 charged  species~\cite{KK,Gudnason:2006ug,Gudnason:2006yj,Barr:1990ca,Nussinov:1985xr,KK2,unesco,iwara}.

 After formation during the Standard Big Bang Nucleosynthesis (BBN) era, $^4He$ screens the
$O^{--}$ charged particles in composite $(^4He^{++}O^{--})$ {\it
$OHe$} ``atoms'' \cite{I}.
In all the models of $OHe$, $O^{--}$ behaves either as a lepton or
as a specific ``heavy quark cluster" with strongly suppressed hadronic
interactions. However $OHe$ is a strongly interactive massive particle (SIMP) due to the $^4He$ nucleus.
The cosmological scenario of the $OHe$ Universe involves only one parameter
of new physics i.e. the mass of O$^{--}$. Such a scenario is insensitive to the properties of $O^{--}$ (except its mass), since the main features of the $OHe$ dark atoms are determined by their nuclear interacting helium shell. In terrestrial matter such dark matter species are slowed down and cannot cause significant nuclear recoil in the underground detectors, making them elusive in direct WIMP search experiments (where detection is based on nuclear recoil) such as CDMS, XENON100 and LUX. The positive results of DAMA and CoGeNT experiments (see \cite{DAMAtalk} for review and references) can find in this scenario a nontrivial explanation due to a low energy radiative capture of $OHe$ by intermediate mass nuclei~\cite{mpla,DMRev,DDMRev}. This can explain the negative results of the XENON100 and LUX experiments. The rate of this capture is
proportional to the temperature: this leads to a suppression of this effect in cryogenic
detectors, such as CDMS. 

OHe collisions in the central part of the Galaxy can potentially lead to OHe
excitations and de-excitations with pair production in E0 transitions that can explain the
excess of the positron-annihilation line, observed by INTEGRAL in the galactic bulge \cite{DMRev,DDMRev,KK2,CKWahe}.

One should note that the nuclear physics of OHe is in the course of development. The basic feature that determines the elastic  OHe-nuclei interactions is an electric dipole potential due to the OHe's charged constituents. This potential requires a specific  quantum mechanical analysis~\cite{CKW}. The lack of a Coulomb barrier seems to lead  inevitably to overproduction of anomalous isotopes \cite{CKW2}. However one should study precisely how  quantum tunneling is modified in this case.

It has been shown \cite{KK,Belotsky:2014,KK2,unesco,iwara} that a two-component dark atom scenario is also possible and can be naturally realized in the framework of  WTC, where both stable doubly charged technileptons $\zeta^{--}$, playing the role of $O^{--}$, and positively doubly charged technibaryons $UU$ are present. 
The two-component dark matter scenario is realised by having an abundance of  $\zeta^{--}$, which binds with $^4He^{++}$, and a much smaller excess of $UU^{++}$, that binds with $\zeta^{--}$, forming a WIMP-like $UU\zeta$ component of dark matter.
 In the framework of WTC, $UU^{++}$ can be metastable, with a dominant decay channel to a pair of positively charged leptons. We argued in \cite{Belotsky:2014} that even a $10^{-6}$ fraction of such positively charged techniparticles with a mass of $\sim 1$ TeV  and a lifetime of~$\sim 10^{20} \s$,  decaying   to $e^+e^+$, $\mu^+ \mu^+$, and $\tau^+ \tau^+$,  can potentially provide an explanation for the observed excess of cosmic ray positrons, while being compatible with the observed gamma-ray background.

The anomalous excess of high-energy positrons in cosmic rays was first observed by PAMELA \cite{PAMELA} and was later confirmed by AMS-02 \cite{AMS-2old}. These results generated widespread interest, since the corresponding effect cannot be explained by positrons of only secondary origin and requires primary positron sources, e.g. annihilations or decays of dark matter particles.
Recently the AMS-02 collaboration has reported new results on positron and electron fluxes in cosmic rays~\cite{AMS2} and positron fraction~\cite{AMS1}. These measurements cover the energy range  $0.5-700$ GeV for electrons and  $0.5-500$ GeV for positrons and provides important information on the origins of primary positrons in the cosmic rays. In particular, the new results show for the first time, that  the positron fraction no longer exhibits an increase with energy above $\sim 200$ GeV.
The possibility of explaining the cosmic positron excess by  decays of $UU$ particles, comprising the tiny WIMP-component of dark matter in the considered scenario, was discussed in detail in~\cite{Belotsky:2014}. Here we make a blind search within the whole parameter space and we find the optimal values that give the best possible fit to both AMS-02 data on cosmic positron flux and FERMI/LAT data on diffuse gamma-ray flux ~\cite{FERMI}.

The paper is organised as follows: in section~\ref{fluxcalc} we briefly discuss the high-energy positron abundance in cosmic rays and the calculation of cosmic positron and gamma-ray fluxes from decays of dark matter particles and in section~\ref{fitting} we search for the best-fit parameter values and discuss the prospects of positron anomaly explanation in the framework of the considering model. We conclude in section~\ref{concl}.

\section{Positron and Photon Flux due to $UU$ decays }\label{fluxcalc}

	The abundance of high-energy positrons in cosmic rays (also known as ``positron anomaly"), which was first observed by HEAT~\cite{HEAT} and was later confirmed by various experiments, such as ATIC~\cite{ATIC}, AMS-01~\cite{AMS-1}, Fermi~\cite{Fermi}, PAMELA~\cite{PAMELA} and AMS-02 \cite{AMS-2old}, can be a possible astrophysical manifestation of  metastable dark matter. Alternative explanations of this phenomenon include old supernova remnants~\cite{Kohri2015} ans pulsars~\cite{Feng2015}. Here we are going to consider the case where an unstable, though long-living, doubly-charged heavy particle $UU^{++}$, which is part of a tiny ($\sim 10^{-6}$) WIMP-like component of dark matter $UU\zeta$, decays as $UU\rightarrow e^+e^+, \mu^+\mu^+, \tau^+\tau^+$ in principle with different relative probabilities. All decay modes produce positrons and photons either directly or through decay cascades. Photons appear in leptonic decays as final state radiation (FSR), arising from $\tau$ decays to hadrons and from inverse Compton scattering (ICS) on the interstellar electromagnetic background. The latter processes are not taken into account in this work, since their contribution was found to be subdominant \cite{Belotsky:2014}. Since there is no production of antiprotons in our scenario, diffuse gamma background gives the most stringent constraints.\\

Here we shall briefly remind the calculations of positron and gamma-ray flux, done in \cite{Belotsky:2014}. The positron flux can be estimated as\footnote{An accurate comparison to the semi-analytical approach, presented in \cite{Ibarra2013}, didn't point out any significant difference.}
\begin{equation}
F_{pos}(E)=\frac{c}{4\pi}\frac{n_{\rm loc}}{\tau}\frac{1}{\beta E^2}
\int_E^{m/2} \frac{dN}{dE_0} Q(\lambda(E_0,E))dE_0.
\label{positronFlux}
\end{equation}
Here $c$ is the speed of light, $\tau$ is the lifetime of $UU$, $n_{\rm loc}=\xi\cdot (0.3\,{\rm  GeV/cm^3})m_{UU}^{-1}$ is the local number density  of $UU$ particles with $\xi=10^{-6}$ being the fraction of dark matter in the form of $UU\zeta$, $m_{UU}$ is the mass of $UU$, $m$ is the mass of the positron, $dN/dE_0$ is the number of positrons produced in a single decay (obtained using Pythia 6.4~\cite{Pythia}), $\beta \sim 10^{-16}$ s$^{-1}$GeV$^{-1}$ is the average energy loss rate and the factor

\begin{equation}
Q=1-\frac{(\lambda-h)^2(2\lambda+4)}{2\lambda^3}\eta(\lambda-h)-\frac{2h(\lambda^2-r^2)}{3\lambda^3}\eta(\lambda-R)
%\label{Qfact}
\end{equation}\\
takes into account effectively the finite size of the magnetic halo ($h \sim 4 \kpc$, $R \sim 15 \kpc$),  $\lambda$ is the characteristic diffusion length, $\eta$ denotes a Heaviside step-function (see~\cite{Belotsky:2014} for details).
Here we will be interested in energies $> 20 \GeV$, where the effects of solar modulation are negligible \cite{AMS2}. The spectrum of secondary positron was taken from~\cite{secspec}.

For gamma rays produced by $UU$ decays in our Galaxy, the flux arriving at the Earth is given by
\begin{equation}
F_{\gamma}^{\rm (G)}(E)=\frac{n_{\rm loc}}{\tau}\frac{1}{4\pi \Delta\Omega_{\rm obs}}\int_{\Delta\Omega_{\rm obs}}\frac{n(r)}{n_{\rm loc}}dld\Omega \cdot \frac{dN_{\gamma}}{dE},
\label{FSRgal}
\end{equation}
where we use an isothermal number density distribution $\frac{n(r)}{n_{\rm loc}}=\frac{(5{\rm \,kpc})^2+(8.5{\rm \,kpc})^2}{(5{\rm \,kpc})^2+r^2}$. Here $l$ and $r$ are respectively the distances from the Earth and the Galactic center. We obtain the averaged flux over the solid angle $\Delta\Omega_{\rm obs}$ corresponding to Galactic lattitudes $|b|>20^\circ$ and Galactic longitude $0<l<360^{\circ}$.
Out of our Galaxy decays of $UU$ are homogeneously distributed over the Universe and should also contribute to the observed gamma-ray flux. This contribution is
\begin{eqnarray}
F_{\gamma}^{\rm (U)}(E)=\frac{c}{4\pi}\frac{\langle n_{\rm mod}\rangle}{\tau}\int \frac{dN}{dE}dt=
\frac{c\langle n_{\rm mod}\rangle}{4\pi\tau}\times\nonumber\\ \times \int_0^{\min(1100,\frac{m}{2E}+1)} \frac{dN}{dE_0}(E_0=E(z+1))\frac{H_{\rm mod}^{-1}dz}{\sqrt{\Omega_{\Lambda}+\Omega_m(z+1)^3}},
%\frac{3\langle n_{\rm mod}\rangle ct_{\rm mod}h}{8\pi\tau}\times\nonumber\\ \times \int_0^{\min(1100,\frac{m}{2E}+1)} \frac{dN}{dE_0}(E_0=E(z+1))\frac{dz}{\sqrt{\Omega_{\Lambda}+\Omega_m(z+1)^3}},
\label{FU}
\end{eqnarray}
where $z=1100$ corresponds to the recombination epoch, $\langle n_{\rm mod}\rangle$ is the current cosmological number density of $UU$, $H_{\rm mod}^{-1}=\frac{3}{2}t_{\rm mod}\sqrt{\Omega_{\Lambda}}\ln^{-1}\left(\frac{1+\sqrt{\Omega_{\Lambda}}}{\sqrt{\Omega_{m}}}\right)$ is the inverse value of the  Hubble parameter with $t_{\rm mod}$ being the  age of the Universe. %$h=\sqrt{\Omega_{\Lambda}}\ln\left(\frac{1+\sqrt{\Omega_{\Lambda}}}{\sqrt{\Omega_{m}}}\right)$,
$\Omega_{\Lambda}$ and $\Omega_m=1-\Omega_{\Lambda}$ are respectively the current vacuum and matter relative densities.	
The total gamma-ray flux arriving at the Earth is 
	\begin{equation}
	F_{\gamma} (E) = F_{\gamma}^{(G)}(E) + F_{\gamma}^{(U)}(E).
	\label{gammaFlux}
	\end{equation}
	In \cite{Belotsky:2014} we showed that the considered model can in principle accommodate nicely the AMS-02 data on positron abundance in cosmic rays \cite{AMS-2old}, being consistent with the FERMI/LAT data on diffuse gamma-ray background (inclusive photon spectrum) \cite{FERMI}. In the following section we find the best fit parameters, taking into account the latest relevant data from AMS-02 \cite{AMS2,AMS1} and FERMI/LAT \cite{FERMIlate}.

	\section{Fitting AMS-02 and FERMI Data}\label{fitting}

	 In order to find the best possible fit for the data we vary four parameters: the mass of the decaying particle $M$, it's lifetime $\tau$ and two values of branching ratios $Br_{ee}$ and $Br_{\mu\mu}$ (the third one is automatically fixed by $Br_{\tau\tau} = 1 - Br_{ee} - Br_{\mu\mu}$). To find the parameter values which provide the best fit of the positron flux data and do not contradict the diffuse gamma-background data, we minimize the following function 
	
	\begin{multline}
	\tilde{\chi}^2 = \frac{1}{N_{d.o.f.}} \left[ \sum_{i=1}^{N_{ams}}\left(\frac{\Phi^{i}_{ams} - F_{pos}(E^{i}_{ams},\vec{a})}{\sigma^{i}_{ams}}\right)^2 + \right.\\ \left. + \sum_{j=1}^{N_{fermi}}\left(\frac{\Phi^{j}_{fermi} - F_{\gamma}(E^{j}_{fermi},\vec{a})}{\sigma^{j}_{fermi}}\right)^2 \cdot \theta \left(F_{\gamma}(E^{j}_{fermi},\vec{a}) - \Phi^{j}_{fermi}\right) \right].
	\label{chi2tilde}
	\end{multline}
	
		\begin{figure}[h!]
			\centering
			\includegraphics[scale = 0.43]{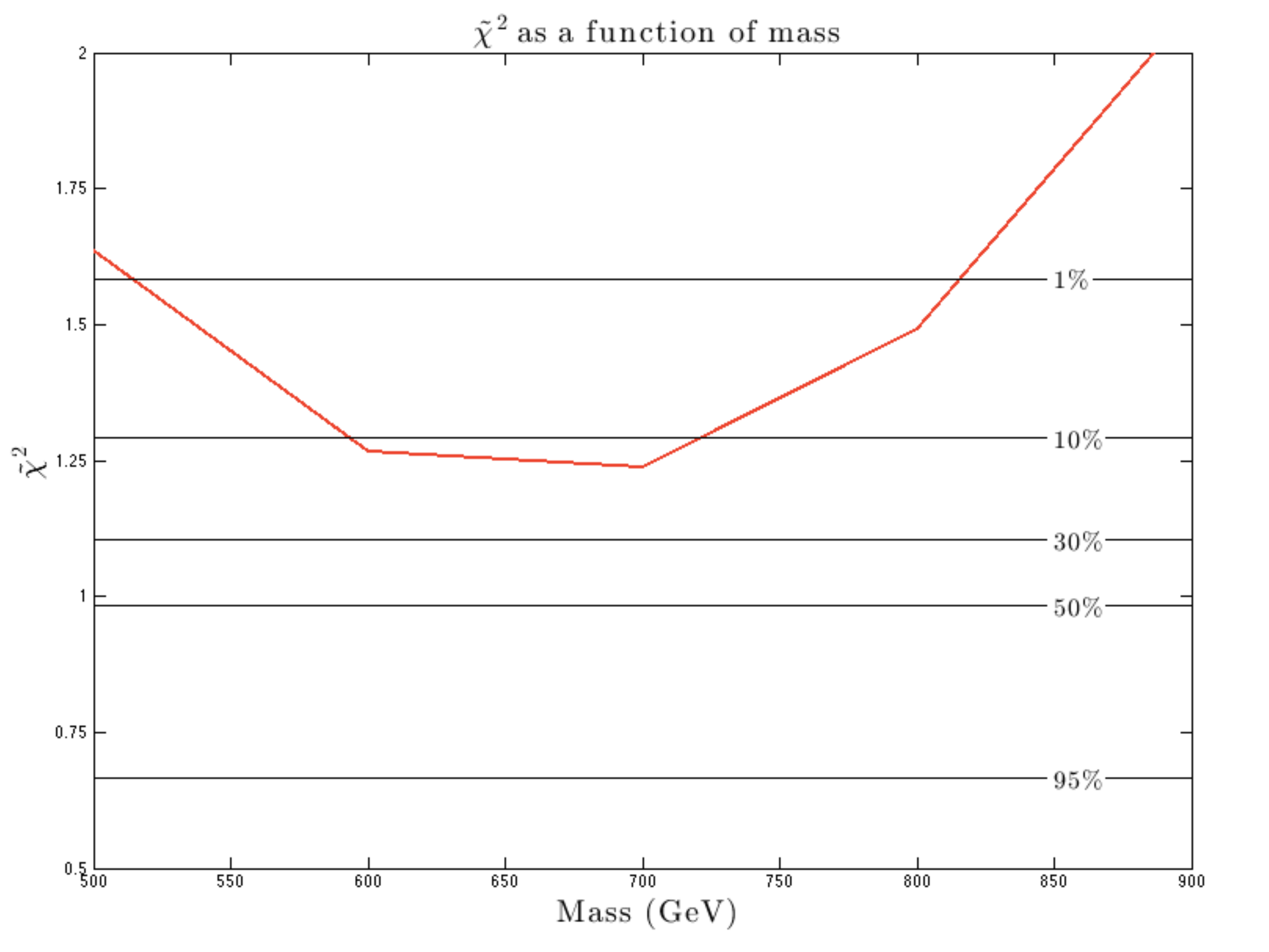}
			\caption{$\tilde{\chi}^2$ as a function of particle mass. For each value of the $UU$ mass we have fixed the residuals parameters to the  values that provide the best fit of the data (see text). Horizontal lines show the p-value for the corresponding $\tilde{\chi}^2$, which can be roughly considered as the probability of the viability of the scenario.}
			\label{chi2new}
		\end{figure}
		
		\begin{figure}[h!]
			\centering
			\includegraphics[scale = 0.38]{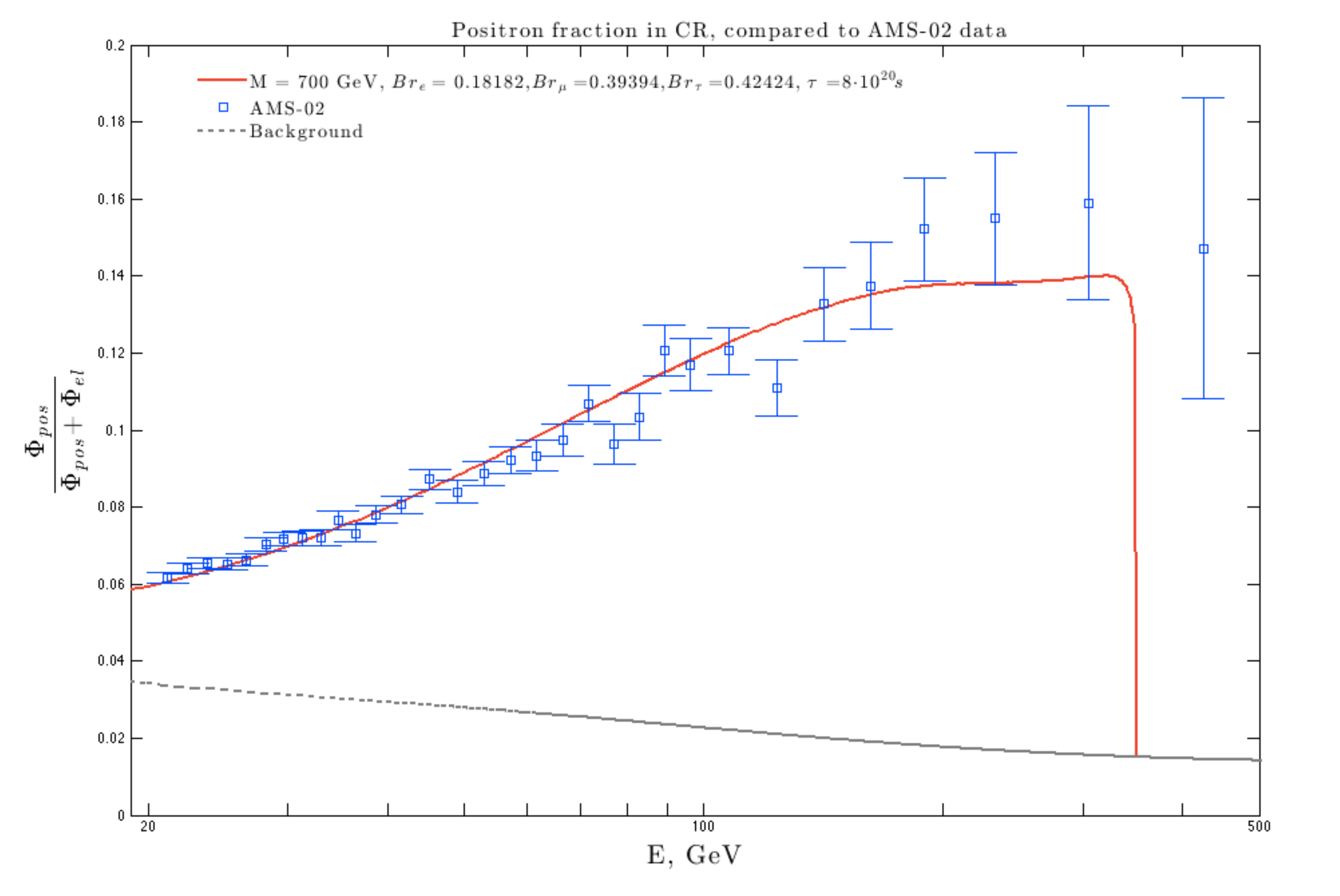}
			\caption{The positron fraction in the cosmic rays from decays of dark matter particles (red curve), corresponding to the best-fit  parameters  $(M = 700 \GeV$, $\tau = 8 \cdot 10^{20} \s, Br_{ee} = 0.182, Br_{\mu\mu} = 0.394, Br_{\tau\tau} = 0.424)$, and the fraction of secondary positrons (gray line), compared to the latest AMS-02 data~\cite{AMS1} (blue dots).}
			\label{positrons}
		\end{figure}
		
		\begin{figure}[h!]
			\centering
			\includegraphics[scale = 0.39]{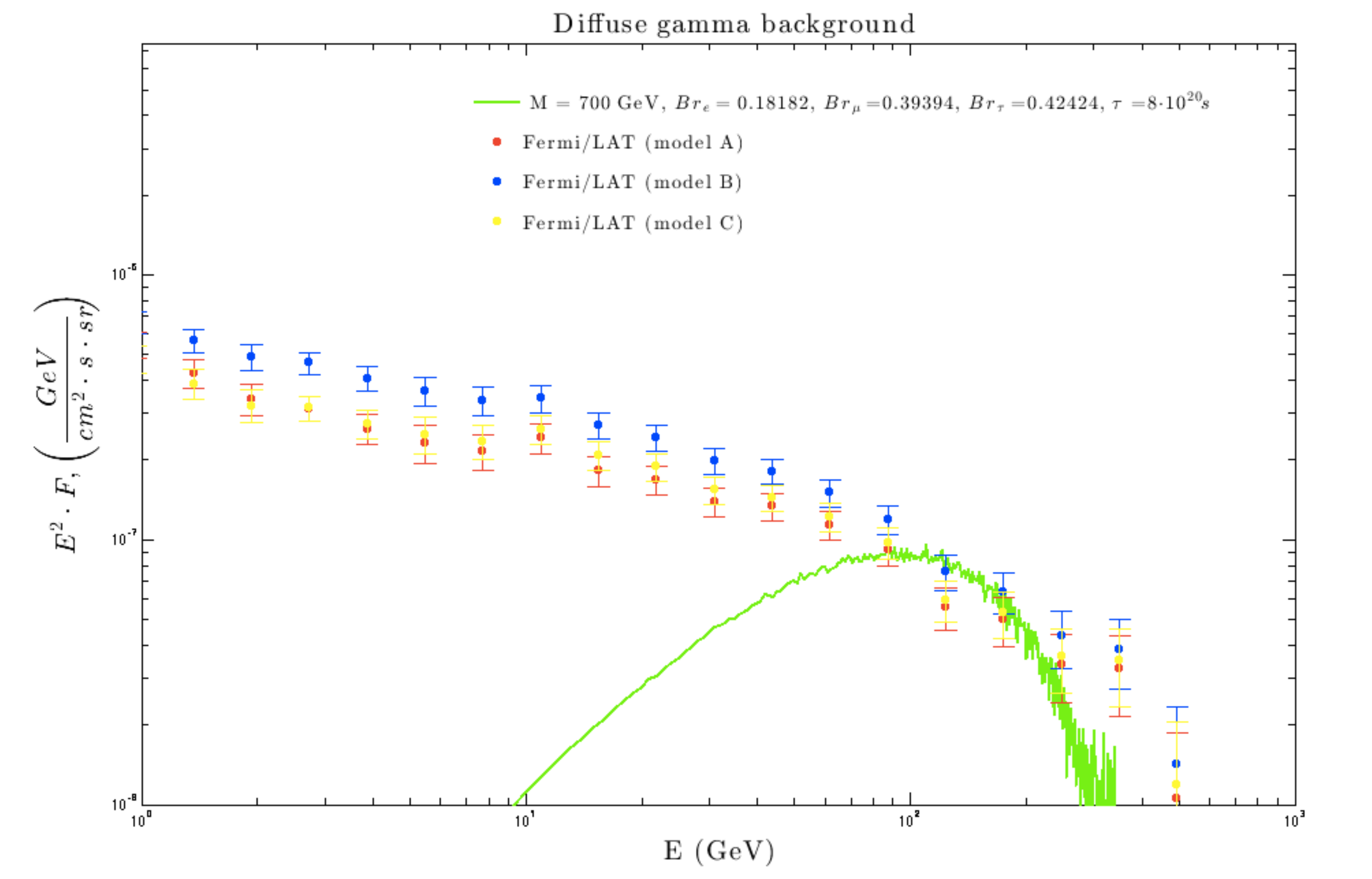}
			\caption{Gamma-ray flux multiplied by $E^2$ from decays of dark matter particles in the Galaxy and beyond (green curve), corresponding to the best-fit parameters $\left(  M = 700 \GeV, \tau = 8 \cdot 10^{20} \s, Br_{ee} = 0.182, Br_{\mu\mu} = 0.394, Br_{\tau\tau} = 0.424\right)$, compared to the latest FERMI/LAT data on isotropic diffuse gamma-ray background \cite{FERMIlate} ($|b| > 20^{\circ}, 0^{\circ} \le l < 360^{\circ}$ with point sources removed and without diffuse emission attributed to the interactions of Galactic cosmic rays with gas and radiation fields (foreground)); here three different foreground models A (red dots), B (blue dots) and C (yellow dots) are shown). In our analysis we have used model B.}
			\label{gammaray}
		\end{figure}
		
		\begin{figure}[h]
			\centering
			\includegraphics[scale = 0.41]{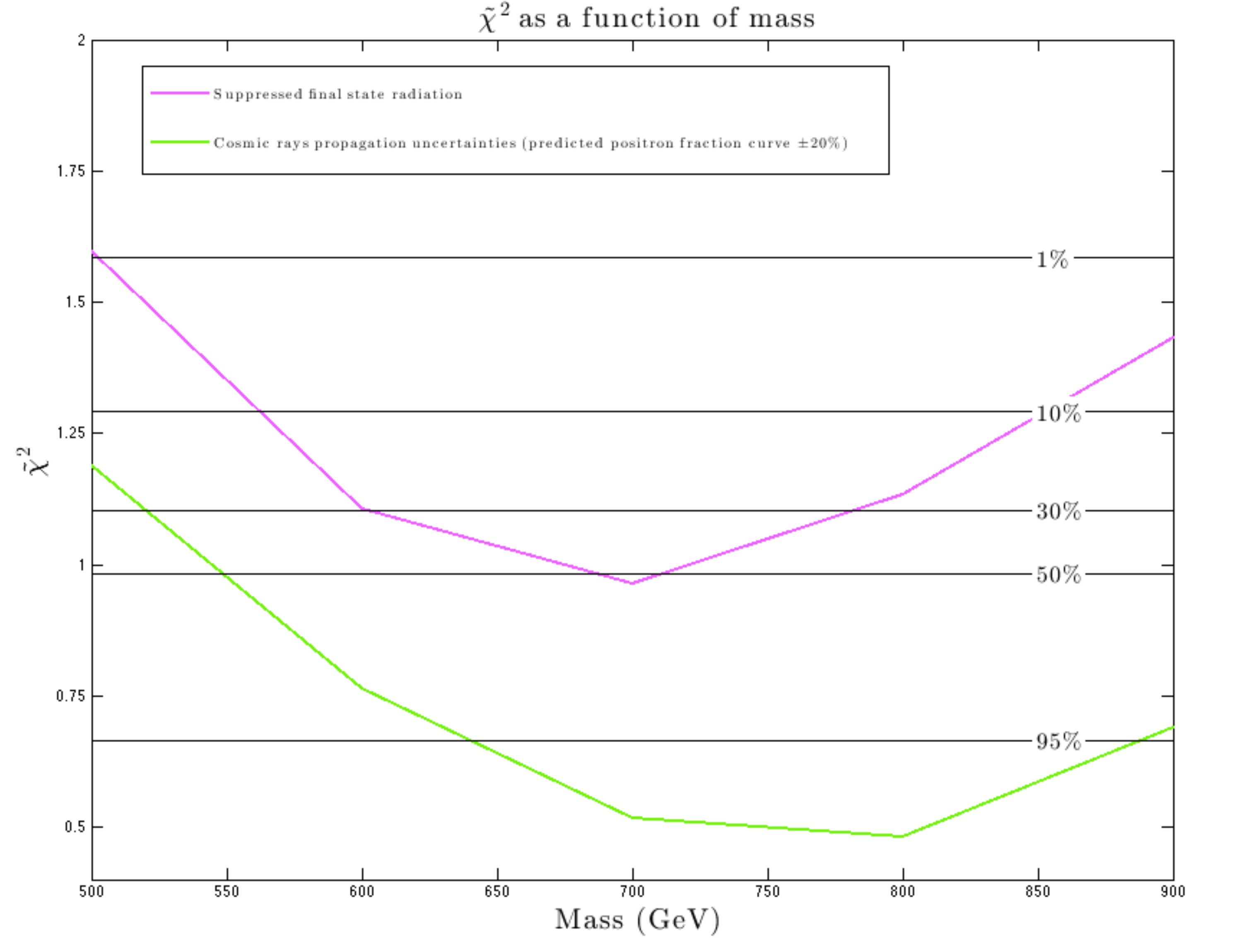}
			\caption{$\tilde{\chi}^2$ as a function of $UU$ mass for the two possibilities  described in the text: the green (lower) curve is obtained under the assumption that the theoretical values of the flux have a 20\% error in all the points (due to the uncertainties in cosmic-ray propagation); the magenta (upper) curve corresponds to the case where FSR is turned off (i.e. the case of two identical particles in the final state, where FSR could be significantly suppressed).}
			
			%i.e. $\sim 20\%$ theoretical uncertainty in the propagation model (green curve) and  suppression of FSR (magenta curve). Horizontal lines show the p-value for the corresponding value of $\tilde{\chi}^2$, which can be roughly considered as the probability of the viability of the scenario.}
			\label{chi2without}
		\end{figure}
		
	Here $(E_{ams}, \Phi_{ams})$ is a set of AMS-02 (energy, positron flux) data points ($N_{ams} = 37$), $F_{pos}(E,\vec{a})$ denotes the predicted positron flux (see Eq.~\ref{positronFlux}), $\sigma_{ams}$ is the AMS-02 statistical error, $(E_{fermi}, \Phi_{fermi})$ is a set of FERMI/LAT IGRB data points ($N_{fermi} = 7$ - we choose only data points in the high-energy region, where the FERMI/LAT data give the most stringent limit on the predicted gamma-ray flux), and $F_{\gamma}(E,\vec{a})$ stands for the predicted gamma-ray flux (see Eq.~\ref{gammaFlux}). %$F_{pos}$ and $F_{\gamma}$ are determined by the same set of parameters. 
	$\sigma_{fermi}$ corresponds to the FERMI/LAT statistical error, and $\theta \left(F_{\gamma}(E_{fermi},\vec{a}) - \Phi_{fermi}\right)$ is a Heaviside step function (it excludes the contribution of the points where $F_{\gamma} < \Phi_{fermi}$). To simplify this analysis we evaluate the minimal value of $\tilde{\chi}^2$ for each value of mass $M$ in the range $500 - 1200 \GeV$ (with a $100 \GeV$ step). Once the mass is fixed, the parameters we scan over are $\vec{a} = \{\tau, Br_{ee}, Br_{\mu\mu}\}$. The number of statistical degrees of freedom is $N_{d.o.f.} = N_{ams} + N_{fermi} - 3 - 1$. The lower limit on the mass value comes from the search for long-lived doubly charged particles at the LHC~\cite{PLB2013}. The rest of the model parameters can have any reasonable value (branching ratios should have a positive value up to 1 and the lifetime should be around $10^{20}$ s). By using the aforementioned method we obtain for each value of $UU$ mass the set of optimal parameters that fit the observational data in the best way. To probe whether or not a particular scenario can provide a good agreement with the experimental data, we treat $\tilde{\chi}^2$ as a common $\chi^2$ statistical test  and  use the Pearson's chi-squared test. 
	
	The results of our analysis are presented in Fig.~\ref{chi2new}. As it can be seen the best possible scenario takes place at $\tilde{\chi}^2 = 1.236$ for $M = 700 \GeV, \tau = 8 \cdot 10^{20} \s, Br_{ee} = 0.182, Br_{\mu\mu} = 0.394, Br_{\tau\tau} = 0.424$. However, it is clear that the obtained parameters provide a not-so-good description of experimental data due to the constraints, following from FERMI/LAT data. For this set of parameters, we show respectively in Fig.~\ref{positrons} and \ref{gammaray}  the predicted positron fraction and gamma-ray background, compared to the experimental data.  

	Despite the fact that the fit is not good,  there are ways that can alter the picture and render the considered scenario viable. 
	One possibility is that FSR production can turn out to be suppressed in the case of two identical particle final state (like $e^+e^+$), compared to the case, where two non-identical particles (like $e^+e^-$) are created, as it was assumed in our calculations.
%	One way is related to a possible suppression of final state radiation in case where two identical particles $(e^+e^+,\mu^+\mu^+ {\rm ~or~} \tau^+\tau^+)$ are created, comparing  (note that in all the calculations presented here we assumed that FSR production is equal to the case, when a lepton-antilepton pair is born). 
A second way out is related to the current uncertainties of the cosmic-ray propagation model, that includes  errors of diffusion coefficients, unaccounted effects such as re-acceleration and choice of source distribution and halo geometry. A rough estimate shows (see Fig.~\ref{chi2without}) that each of the two proposed ideas applied to the considered model, can significantly reduce the $\tilde{\chi}^2$ value for a considerable range of $UU$ masses. However, a more detailed study and better understanding of the uncertainties of cosmic rays propagation is required in order to clarify the viability of the considered $UU$ decay model. In addition, one should note that  inverse Compton scattering processes were not taken into account and their contribution into diffuse gamma-ray flux can also affect slightly the fit. 
	
\section{Conclusions}\label{concl}
If the Standard Model can serve as a guiding theory for the explanation of the dark matter puzzle, there is no reason, why dark matter must be a single particle~\cite{DMRev}. Multi-component dark matter with  components that contribute different amounts to the total dark matter abundance is a possible scenario. There can easily be a dark matter realization where a dominant SIMP-like component like $He\zeta$ is accompanied by a small WIMP-like component $UU\zeta$. This can be particularly interesting because there have been several issues with the validity of the Cold Collisionless Dark Matter paradigm recently. The core vs cusp problem in dwarf galaxies, the number of satellite galaxies and the ``too big to fail problem" might be manifestations of the fact that dark matter might be warmer  than what was thought until now (see e.g.~\cite{Petraki:2014uza} and references therein). A two component dark matter scenario can in fact resolve the aforementioned issues. The scenario considered here involves a form of neutral $OHe$ dark atoms made of stable heavy doubly charged particles and primordial He nuclei bound by ordinary Coulomb interactions. This scenario can be realized in the framework of Minimal Walking Technicolor, where an exact relation between the dark matter density and baryon asymmetry can be naturally obtained. Strict conservation of technilepton charge together with approximate conservation of technibaryon charge results in the prediction of two types of doubly charged species with strongly unequal excess, i.e. dominant negatively charged technileptons $\zeta^{--}$ and a strongly subdominant component of technibaryons $UU^{++}$, bound with $\zeta^{--}$ in
  a sparse component of WIMP-like dark atoms ($\zeta^{--}UU^{++}$). Direct searches for WIMPs put severe constraints on the presence of this component. However we have demonstrated in~\cite{Belotsky:2014} that
 the existence of a metastable positively doubly charged techniparticle, forming this tiny subdominant WIMP-like dark atom component and satisfying the direct WIMP searches constraints, can play an important role in the indirect effects of dark matter. We found that decays of such positively charged constituents of WIMP-like dark atoms to  leptons $e^+e^+, \mu^+\mu^+,\tau^+ \tau^+$ can explain the observed excess of high energy cosmic ray positrons, while being compatible with the observed gamma-ray background. These decays are naturally  facilitated by GUT scale interactions. The best fit of the data takes place for a mass of this doubly charged particle around 1 TeV or below making it accessible in the current run of LHC. Our refined best fit analysis of the recent AMS-02  and FERMI/LAT data presented here, can provide a crucial test for the decaying dark atom hypothesis especially since LHC in its  second run will be able to probe the parameter space where this dark atom model is viable.
 
\section*{Acknowledgements}
 The work by M.Kh. and M.L. on initial cosmological conditions was supported by the Ministry of Education and Science of Russian Federation, project 3.472.2014/K  and the work of K.B., M.Kh. and M.L. on the forms of dark matter was supported by grant RFBR 14-22-03048. C.K. is supported by the Danish National Research Foundation, Grant No. DNRF90.

%% The bibliography section

\end{document}